\documentclass[journal]{IEEEtran}
\usepackage{multirow}
\usepackage{graphicx}
\usepackage{amsmath}
\usepackage{color}
\usepackage{epsfig}
\usepackage{amsfonts}
\usepackage{amssymb}
\usepackage{amsthm}
\usepackage[usenames,dvipsnames]{pstricks}
\usepackage{epstopdf}
\usepackage{algorithm}
\usepackage{caption}
\usepackage{subcaption}
\usepackage{mathtools}
\usepackage[noend]{algpseudocode}
\usepackage{comment}

\begin{document}
\title{\huge{{\color{black}
Input-Output Relation and Performance of RIS-Aided OTFS with Fractional Delay-Doppler} }}
\author{Vighnesh S Bhat, Gandhodi Harshavardhan, and A. Chockalingam \\
Department of ECE, Indian Institute of Science,
Bangalore 560012
\thanks{Copyright (c) 2019 IEEE. Personal use of this material is permitted. Permission from IEEE must be obtained for all other uses, in any current or future media, including reprinting/republishing this material for advertising or promotional purposes, creating new collective works, for resale or redistribution to servers or lists, or reuse of any copyrighted component of this work in other works.}
\vspace{-4mm}
} 

\maketitle

\begin{abstract}
Reconfigurable intelligent surfaces (RIS) and orthogonal time-frequency space (OTFS) modulation have gained attention in recent wireless research. RIS technology aids communication by reflecting the incident electromagnetic waves towards the receiver, and OTFS modulation is effective in high-Doppler channels. This paper presents an early investigation of RIS-aided OTFS in high-Doppler channels. We derive the end-to-end delay-Doppler (DD) domain input-output relation of a RIS-aided OTFS system, considering rectangular pulses and fractional delay-Doppler values. We also consider a Zak receiver for RIS-aided OTFS that converts the received time-domain signal to DD domain in one step using Zak transform, and derive its end-to-end input-output relation. Our simulation results show that $i)$ RIS-aided OTFS performs better than OTFS without RIS, $ii)$ Zak receiver performs better than a two-step receiver, and $iii)$ RIS-aided OTFS achieves superior performance compared to RIS-aided OFDM. 
\end{abstract}


\section{Introduction}
\label{sec1}
Next generation mobile communication systems demand improved performance requirements, including enhanced data rate, increased power/spectral efficiency, high reliability, low latency, and high mobility. In the recent literature, reconfigurable intelligent surface (RIS) \cite{b4},\cite{b2}, and orthogonal time-frequency space (OTFS) modulation \cite{b6},\cite{b7},\cite{b11} have emerged as promising physical layer techniques to address such requirements. RIS technology is a power-efficient form of wireless communication, and OTFS technology is a means for providing reliable high-mobility support. RIS aids communication between the transmitter and receiver by smartly controlling the propagation environment with tunable reflecting elements. The reflecting elements dynamically alter the reflection characteristics of the incident electromagnetic wave such that the desired parameters at the receiver are optimized. RIS can also degrade unintended users' signal characteristics, enhancing security/privacy. Since RIS does not require any dedicated energy source and the reflecting elements are passive, the use of RIS makes communication power efficient \cite{ris1}. 

Communication in high-mobility scenarios is a key research topic in next-generation mobile communications. Because of high mobility, the channel experiences rapid variations in time, and existing multicarrier modulation schemes, such as orthogonal frequency division multiplexing (OFDM), are prone to inter-carrier interference, that leads to degraded performance. The newly introduced OTFS modulation \cite{b6} performs significantly better than OFDM in high-mobility environments. OTFS exploits the idea of multiplexing information symbols in delay-Doppler (DD) domain instead of time-frequency (TF) domain. Also, OTFS views time-varying channels in the DD domain where the time variations are slow, which simplifies channel estimation.

Recent studies have investigated the use of RIS in OFDM systems, showing that RIS can enhance OFDM performance \cite{b5}-\cite{RIS_OFDM_3}. Most works on RIS mainly consider block-wise quasi-static fading. However, wireless channels are time-varying due to user mobility. Also, next-generation wireless systems will need to provide ubiquitous tetherless connectivity in high-mobility scenarios with increased power/spectral efficiency. Therefore, a combination of RIS and OTFS can simultaneously offer the benefits of both power efficiency (due to RIS) and robustness in high mobility (due to OTFS). Recently, \cite{RIS_OTFS} considered a RIS-aided OTFS system, derived its input-output relation, and showed performance gains. While this study showed the benefit of using RIS in OTFS, it considered ideal bi-orthogonal pulses and integer DD values. On the other hand, practical pulse shapes do not obey bi-orthogonality condition, and fractional DD values will be encountered in practice. This paper focuses on RIS-aided OTFS with rectangular pulses and fractional DD values, which are practically more relevant. The contributions in this paper are summarized as follows.
\begin{itemize}
\item We derive the end-to-end DD domain input-output relation of a RIS-aided OTFS system with rectangular pulses at the transmitter and receiver and fractional DD values. We carry out this derivation for two types of OTFS receivers, namely, 1) a {\em two-step receiver}, where the received time domain (TD) signal is converted into a DD domain signal in two steps, viz., TD to time-frequency (TF) domain using Wigner transform and TF domain to DD domain using symplectic finite Fourier transform, and 2) a {\em single-step Zak receiver}, which uses Zak transform to directly convert the received TD domain signal to DD domain. The derived input-output relations can aid  transceiver algorithms development for RIS-aided OTFS and performance evaluation.
\item Our simulation results show that 1) RIS improves the performance of OTFS, 2) 
single-step Zak receiver performs better than two-step receiver, and 3) RIS-OTFS achieves superior performance compared to RIS-aided OFDM.
\end{itemize}

The rest of the paper is organized as follows. The input-output relation of RIS-aided OTFS with two-step receiver is presented in Sec. \ref{sec2}. The input-output relation of RIS-aided OTFS with single-step Zak receiver is presented in Sec. \ref{sec3}. Results and discussions are presented in \ref{sec4}. Conclusions are presented in Sec. \ref{sec5}.

\section{RIS-aided OTFS with two-step receiver}
\label{sec2}
\begin{figure}[t]
\centering
\includegraphics[width=8.5cm,height=4.5cm]{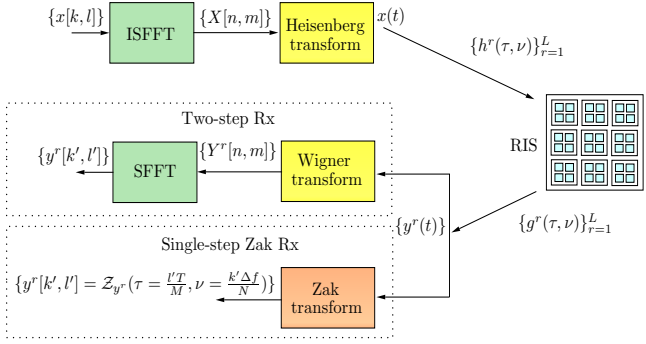}
\caption{RIS-aided OTFS system.}
\label{RIS_two_step}
\vspace{-2mm}
\end{figure}

A RIS-aided OTFS system consists of an OTFS transmitter, an OTFS receiver, and a RIS as shown in Fig. \ref{RIS_two_step}. The RIS has $K$ reflecting elements, whose phases can be controlled dynamically to enhance desired parameters at the receiver. The adjacent elements with highly correlated channel gains are grouped to a sub-surface to reduce the overhead in the reflection phase design. Therefore, RIS has $L$ sub-surfaces, where each sub-surface consists of $N_s=K/L$ adjacent elements. Let the reflection coefficient at the $r$th sub-surface be defined as $\phi_r=\gamma_r e^{j\theta_r}$, $r=1,\cdots,L$, where $\gamma_r\in [0,1]$, $\theta_r\in[-\pi,\pi]$ are the reflection amplitude and phase of the $r$th sub-surface, respectively. 

Information symbols $x[k,l]$ from a modulation alphabet ${\mathbb A}$, $k=0,\cdots,N-1, l=0,\cdots,M-1$ are multiplexed in the DD domain at the OTFS transmitter, where $N$ and $M$ are the number of Doppler and delay bins, respectively, in the DD grid. The OTFS transmitter operations involve inverse symplectic finite Fourier transform (ISFFT) and Heisenberg transform to convert a DD domain signal to TF domain signal and TF signal to TD signal, respectively.
The TF domain signal obtained through ISFFT is given by 
\begin{align}
X[n,m]=\frac{1}{MN}\displaystyle \sum_{k=0}^{N-1}\displaystyle \sum_{l=0}^{M-1}x[k,l]e^{j2\pi(\frac{nk}{N}-\frac{ml}{M})},
\label{x_TF}
\end{align}
$n=0\cdots N-1$ and $m=0\cdots M-1$ are the indices of the time and frequency bins in the TF grid, respectively, and the TD signal obtained through Heisenberg transform is given by 
\begin{align}
x(t)=\displaystyle \sum_{n=0}^{N-1}\displaystyle \sum_{m=0}^{M-1}X[n,m]g_{tx}(t-nT)e^{j2\pi m\Delta f(t-nT)},
\label{x_time}
\end{align}
where $g_{tx}(.)$ is the transmit pulse. Each OTFS frame has $NT$ duration and occupies $M\Delta f$ bandwidth with $\Delta f=1/T$. The received TD signal at $r$th sub-surface of RIS, denoted by $z^r(t)$, is given by \cite{ZAK3}
\begin{align}
z^r(t)=\displaystyle \sum_{p=1}^{P_1}h_p^rx(t-\tau_p^{r,1})e^{j2\pi\nu_p^{r,1}(t-\tau_p^{r,1})},
\end{align}
where $P_1$ is the number of paths in the transmitter-to-RIS link (i.e., 1st link), $h_p^r$, $\tau_p^{r,1}$, and $\nu_p^{r,1}$ are the channel gain, delay, and Doppler of the $p$th path associated with $r$th sub-surface in the 1st link, respectively. The delays and Dopplers are considered to be fractional, i.e., $\tau_p^{r,1}\overset{\Delta}{=} \frac{\alpha_p^{r,1}+a_p^{r,1}}{M\Delta f}$, $\nu_p^{r,1}\overset{\Delta}{=} \frac{\beta_p^{r,1}+b_p^{r,1}}{NT}$, where
$\alpha_p^{r,1}$, $\beta_p^{r,1}$ are the integers denoting the delay and Doppler indices, respectively, and $a_p^{r,1}$, $b_p^{r,1} \in [-0.5,0.5]$ are the corresponding fractional parts. Similarly, the received TD signal at the receiver associated with $r$th sub-surface, denoted by $y^r(t)$ is given by 
\begin{align}
 y^r(t)=\phi_r\displaystyle \sum_{q=1}^{P_2}g_q^rz^{r}(t-\tau_q^{r,2})e^{j2\pi\nu_q^{r,2}(t-\tau_q^{r,2})},
 \label{rx}
\end{align}
where $P_2$ is the number of paths in the RIS-to-receiver link (i.e., 2nd link), $g_q^r$, $\tau_q^{r,2}$, and $\nu_q^{r,2}$ are the channel gain, delay, and Doppler of the $q$th path associated with $r$th sub-surface in the 2nd link, respectively. The number of paths $P_1,P_2$ depend on the number of dominant reflectors in the environment, and typical values of $P_1,P_2$ in standards range from 4 to 10 \cite{itu}. The total number of channel coefficients of interest, therefore, is $P_1P_2L$.
Substituting for $z^r(t)$, (\ref{rx}) can be written as 
\begin{eqnarray}
y^r(t)\hspace{-1mm} & \hspace{-2mm}=& \hspace{-2.5mm}\phi_r \hspace{-1mm} \displaystyle \sum_{q=1}^{P_2}\sum_{p=1}^{P_1}g_q^r h_p^rx(t-\tau_p^{r,1}-\tau_q^{r,2})e^{j2\pi\nu_p^{r,1}(t-\tau_p^{r,1}-\tau_q^{r,2})} \nonumber \\
& \hspace{-4mm} & \hspace{2mm} e^{j2\pi\nu_q^{r,2}(t-\tau_q^{r,2})}.
\label{rx1_DD}
\end{eqnarray}
In (\ref{rx1_DD}), multiplying and dividing by $e^{j2\pi \nu^{r,2}_q \tau^{r,1}_p}$, we can write
\begin{align}
y^r(t)=\phi_r\sum_{q=1}^{P_2}\sum_{p=1}^{P_1}g_q^rh_p^r \rho^r_{pq}x(t-\tau^r_{pq})e^{j2\pi\nu^r_{pq}(t-\tau^r_{pq})}.
\label{rx2_DD}
\end{align}
where $\rho^r_{pq}\overset{\Delta}{=}e^{j2\pi\nu^{r,2}_q\tau^{r,1}_p}$, $\tau^r_{pq}\overset{\Delta}{=}\tau^{r,1}_{p}+\tau^{r,2}_{q}$, $0\leq\tau^r_{pq}<T$ and $\nu^r_{pq}\overset{\Delta}{=}\nu^{r,1}_{p}+\nu^{r,2}_{q}$.
The received discrete TF signal is obtained through Wigner transform of the TD signal, which is given by 
\begin{eqnarray}
Y^r[n,m]\overset{\Delta}{=}\int_{-\infty}^{\infty}g_{rx}^*(t-nT)y^r(t)e^{-j2\pi m\Delta f(t-nT)}dt,
 \label{y_TF1}
\end{eqnarray}
$m=0,\cdots,M-1, n=0,\cdots,N-1$, where $g_{rx}(.)$ is the receive pulse.
Substituting for $y^r(t)$ from (\ref{rx2_DD}) in (\ref{y_TF1}), we get
\begin{eqnarray}
\hspace{-1mm}
Y^r[n,m]& \hspace{-2mm}=& \hspace{-3mm}\int_{-\infty}^{\infty} \hspace{-2mm} g_{rx}^*(t-nT)\bigg\{\phi_r \sum_{q=1}^{P_2}\sum_{p=1}^{P_1}g_q^r h_p^r \rho^r_{pq}x(t-\tau^r_{pq})\nonumber\\
& \hspace{-2mm} & \hspace{-2mm} e^{j2\pi\nu^r_{pq}(t-\tau^r_{pq})}\bigg\}e^{-j2\pi m\Delta f(t-nT)}dt.
\label{y_TF2}
\end{eqnarray}
Substituting for $x(t)$ from (\ref{x_time}) in (\ref{y_TF2}), we obtain
\begin{eqnarray}
Y^r[n,m] & \hspace{-3mm} = & \hspace{-3mm} \phi_r \sum_{q=1}^{P_2}\sum_{p=1}^{P_1} g_q^r h_p^r \rho^r_{pq}\int_{-\infty}^{\infty}g_{rx}^*(t-nT) \nonumber \\ & \hspace{-15mm} & \hspace{-20mm}  \bigg\{\sum_{n'=0}^{N-1}\displaystyle \sum_{m'=0}^{M-1} \hspace{-2mm}
X[n',m'] g_{tx}(t-\tau^r_{pq}-n'T)
e^{j2\pi m'\Delta f(t-\tau^r_{pq}-n'T)}\bigg\} \nonumber \\ & \hspace{-10mm} & \hspace{-10mm} 
e^{j2\pi\nu^r_{pq}(t-\tau^r_{pq})} e^{-j2\pi m\Delta f(t-nT)}dt. 
\label{y_TF3}
\end{eqnarray}
Rearranging the terms in (\ref{y_TF3}), we write
\begin{eqnarray}
Y^r[n,m]& \hspace{-2mm}=& \hspace{-2mm}\phi_r\sum_{q=1}^{P_2} \sum_{p=1}^{P_1}g_q^r h_p^r \rho^r_{pq}\sum_{n'=0}^{N-1} \sum_{m'=0}^{M-1}X[n',m']\nonumber\\
& \hspace{-20mm} & \hspace{-20mm}
\bigg\{\hspace{-1mm} \int_{-\infty}^{\infty} \hspace{-3mm}  g_{rx}^*(t-nT) g_{tx}(t-\tau^r_{pq}-n'T)e^{j2\pi((m'-m)\Delta f+\nu^r_{pq})t}dt\bigg\}\nonumber\\
& \hspace{-10mm} & \hspace{-10mm}
e^{-j2\pi(m'\Delta f+\nu^r_{pq})\tau^r_{pq}}.
\label{y_TF4}
\end{eqnarray}
The DD domain signal is obtained by applying SFFT on (\ref{y_TF4}), as
\begin{eqnarray}
y^r[k',l']\overset{\Delta}{=}\sum_{n=0}^{N-1} \sum_{m=0}^{M-1}Y^r[n,m]e^{j2\pi(\frac{ml'}{M}-\frac{nk'}{N})}.
\label{y_DD1}
\end{eqnarray}
Let $g_{rx}(t)=g_{tx}(t)=g(t)$ be the rectangular pulse, given by
\begin{align}
    g(t)=\left\{\begin{matrix}
\frac{1}{\sqrt T}, & 0\leq t <T\\ 
 0, & \text{otherwise}
\end{matrix}.\right.
\end{align}
Substituting for $X[n',m']$ from (\ref{x_TF}) and evaluating the integration in (\ref{y_TF4}), we get
\begin{eqnarray}
y[k',l']\hspace{-0.5mm}=\hspace{-0.5mm}\phi_r\hspace{-2mm}\sum_{k=0}^{N-1} \sum_{l=0}^{M-1}\hspace{-1mm} x[k,l] \bigg(\hat{h}^r_1[k',l',k,l]+\hat{h}^r_2[k',l',k,l]\bigg),
\label{y_DD2}
\end{eqnarray}
where 
\begin{eqnarray}
\hat{h}^r_1[k',l',k,l]& \hspace{-2mm}=& \hspace{-2mm} \sum_{q=1}^{P_2} \sum_{p=1}^{P_1}g_q^rh_p^r \rho^r_{pq}e^{-j2\pi\nu^r_{pq}\tau^r_{pq}}\frac{\bigg(1-\frac{\tau^r_{pq}}{T}\bigg)}{M}\nonumber\\
& & 
\Bigg[\frac{1}{N}\sum_{n=0}^{N-1}\displaystyle e^{-j2\pi n\left(\frac{k'-k}{N}-\frac{\nu^r_{pq}}{\Delta f}\right)}\Bigg] \nonumber \\
& & 
 \sum_{m=0}^{M-1} \sum_{m'=0}^{M-1}e^{j\pi\left(1+\frac{\tau^r_{pq}}{T}\right)\left((m'-m)+\frac{\nu^r_{pq}}{\Delta f}\right)}\nonumber\\
& &
e^{j2\pi\left(\frac{ml'}{M}-\frac{m'l}{M}-\frac{m'\tau^r_{pq}}{T}\right)} \nonumber \\
& &
\text{sinc}\bigg(\left((m'-m)+\frac{\nu^r_{pq}}{\Delta f}\right)\hspace{-1mm}\left(1-\frac{\tau^r_{pq}}{T}\right)\bigg), \nonumber
\end{eqnarray} 
and
\begin{eqnarray}
\hat{h}^r_2[k',l',k,l]& \hspace{-2mm}=& \hspace{-2mm} \sum_{q=1}^{P_2} \sum_{p=1}^{P_1}g_q^rh_p^r \rho^r_{pq}e^{-j2\pi\nu^r_{pq}\tau^r_{pq}}e^{-j2\pi\frac{k}{N}}\frac{\left(\frac{\tau^r_{pq}}{T}\right)}{M}
\nonumber\\
& &
\Bigg[\frac{1}{N}\sum_{n=0}^{N-1} e^{-j2\pi n\left(\frac{k'-k}{N}-\frac{\nu^r_{pq}}{\Delta f}\right)}\Bigg] \nonumber \\
& &
 \sum_{m=0}^{M-1} \sum_{m'=0}^{M-1}e^{j2\pi\left(\frac{ml'}{M}-\frac{m'l}{M}-\frac{m'\tau^r_{pq}}{T}\right)}\nonumber\\
& &
e^{j\pi\left(\frac{\tau^r_{pq}}{T}\right)\left((m'-m)+\frac{\nu^r_{pq}}{\Delta f}\right)} \nonumber \\
& &
\text{sinc}\bigg(\left((m'-m)+\frac{\nu^r_{pq}}{\Delta f}\right)\left(\frac{\tau^r_{pq}}{T}\right)\bigg). \nonumber
\end{eqnarray}
Vectorizing (\ref{y_DD2}), we can write
\begin{align}
    \mathbf{y}^r=\phi_r \mathbf{H}^r\mathbf{x},
    \label{vec1}
\end{align}
where $\mathbf{H}^r \in \mathbb{C}^{MN \times MN}$ is the effective cascaded channel matrix for the $r$th sub-surface with the element in its $(l'+k'M+1)$th row and $(l+kM+1)$th column being $\hat{h}_1^r[k',l',k,l]+\hat{h}_2^r[k',l',k,l]$. Finally, the overall end-to-end input-output relation is obtained by adding the reflected signals from all the sub-surfaces at the receiver, as
\begin{eqnarray}
\mathbf{y}& \hspace{-2mm} = & \hspace{-2mm} \sum_{r=1}^{L}\phi_r \mathbf{H}^r \mathbf{x}+\mathbf{n}
\ = \ \mathbf{H}_{\text{eff}}\mathbf{x}+\mathbf{n},
    \label{vec2}
\end{eqnarray}
where $\mathbf{H}_{\text{eff}}=\sum_{r=1}^{L}\phi_r \mathbf{H}^r$, $\mathbf{y}$ is the combined received signal vector from all the sub-surfaces, and $\mathbf{n}$ is the additive noise vector at the receiver.

\section{Single-step Zak receiver for RIS-aided OTFS}
\label{sec3}
In this section, we derive the end-to-end input-output relation for the single-step Zak receiver. A motivation for the combination of ISFFT-based transmitter and Zak-based receiver is as follows. At the transmitter side, the ISFFT-based approach can retain the advantage of building the OTFS transmitter as an overlay on existing multicarrier transmitters. On the other hand, at the user end, the receiver can optionally be implemented using either the SFFT approach or the Zak approach, and therefore a comparison between them is of interest.
The Zak representation of a signal $s(t)$ is defined as \cite{ZAK1},\cite{ZAK2} 
\begin{align}
\hspace{-3mm}
\mathcal{Z}_{s}(\tau,\nu) \overset{\Delta}{=} \sqrt T \sum_{k=-\infty}^{\infty} s(\tau+kT) e^{-j2\pi k\nu T}, \hspace{1mm} \tau,\nu \in (-\infty,\infty).
\label{zak1}
\end{align}
The Zak representation of the received signal $y^r(t)$ is
\begin{align}
\mathcal{Z}_{y^r}(\tau,\nu) = \sqrt T  \sum_{k=-\infty}^{\infty} y^r(\tau+kT) e^{-j2\pi k\nu T}.
\label{zak2}
\end{align}
Substituting for $y^r(t)$, (\ref{zak2}) can be written as
\begin{eqnarray}
\mathcal{Z}_{y^r}(\tau,\nu) & \hspace{-2mm}= & \hspace{-2mm}\sqrt T \phi_r \sum_{k=-\infty}^{\infty} \bigg\{  \sum_{q=1}^{P_2}\sum_{p=1}^{P_1}\rho^r_{pq}g^r_qh^r_p x(\tau-\tau^r_{pq}+kT) \nonumber \\ & \hspace{-2mm} & \hspace{-2mm} e^{j2 \pi \nu^r_{pq}(\tau-\tau^r_{pq}+kT)}\bigg \}e^{-j2\pi k\nu T},
\label{zak3}
\end{eqnarray}
which can be further written as
\begin{eqnarray}
\mathcal{Z}_{y^r}(\tau,\nu) & \hspace{-2mm}= & \hspace{-2mm}  \phi_r \sum_{q=1}^{P_2}\sum_{p=1}^{P_1}\rho^r_{pq}g^r_qh^r_p e^{j2\pi \nu^r_{pq}(\tau-\tau^r_{pq})} \nonumber \\ & \hspace{-2mm} & \hspace{-2mm} \sqrt T\sum_{k=-\infty}^{\infty}    x(\tau-\tau^r_{pq}+kT) e^{-j2\pi(\nu-\nu^r_{pq})kT} \nonumber \\ 
& \hspace{-2mm} = & \hspace{-2mm}
\phi_r \hspace{-0mm} \sum_{q=1}^{P_2}\sum_{p=1}^{P_1} \hspace{-0mm} \rho^r_{pq}g^r_qh^r_p e^{j2\pi \nu^r_{pq}(\tau-\tau^r_{pq})} \nonumber \\
& \hspace{-2mm} & \hspace{-2mm} 
\mathcal{Z}_{x}(\tau-\tau^r_{pq},\nu-\nu^r_{pq}),
\label{zak4}
\end{eqnarray}
where $\mathcal{Z}_x(\cdot)$ denotes the Zak transform of $x(t)$ with delay and Doppler shift of $\tau^r_{pq}$ and $\nu^r_{pq}$ along the delay and Doppler axis, respectively. The Zak transform of the rectangular pulse $g(t)$ is given by \cite{ZAK1}
\begin{align}
    \mathcal{Z}_{g}(\tau,\nu)=e^{j2\pi\nu \left \lfloor \frac{\tau}{T} \right \rfloor T},\hspace{1mm}\tau, \nu,\in (-\infty,\infty).
    \label{zak5}
\end{align}
where $\lfloor \cdot \rfloor$ denotes the floor function. The Zak transform of the transmitted OTFS signal $x(t)$ in (\ref{x_time}) is given by \cite{ZAK1}
\begin{align}
{\mathcal Z}_x(\tau, \nu) & = \sum \limits _{k=0}^{N-1}\sum \limits _{l=0}^{M-1} x[k,l] \, \Psi _{k,l}(\tau,\nu) \,,\, \nonumber \\ \Psi _{k,l}(\tau,\nu) & \stackrel{\Delta }{=} \frac{{\mathcal Z}_g(\tau,\nu)}{MN} \sum \limits _{m=0}^{M-1}\sum \limits _{n=0}^{N-1} \left[ e^{-j 2 \pi n T \left(\nu - k \frac{\Delta f}{N} \right)} \right. \nonumber \\ &\qquad\qquad\qquad\qquad\quad \left. e^{j 2 \pi m \Delta f \left(\tau - \frac{lT}{M} \right)} \right],
\label{zak6} 
\end{align}
where $\mathcal{Z}_g(\tau,\nu)$ is the Zak transform of $g(t)$ defined in (\ref{zak5}). In (\ref{zak6}), $\Psi _{k,l}(\tau,\nu)$ can be viewed as basis signals in the DD domain. Substituting (\ref{zak5}) and (\ref{zak6}) in (\ref{zak4}), we get
\begin{eqnarray}
\mathcal{Z}_{y^r}(\tau,\nu)& \hspace{-2mm}=& \hspace{-2mm} \phi_r\sum_{k=0}^{N-1}\sum_{l=0}^{M-1}x[k,l] \Bigg[ \sum_{q=1}^{P_2} \sum_{p=1}^{P_1} \rho^r_{pq}g^r_qh^r_p \nonumber \\ & \hspace{-20mm} & \hspace{-20mm}  e^{j2\pi\frac{(\nu-\nu^r_{pq})}{\Delta f}\left \lfloor \frac{\tau-\tau^r_{pq}}{T} \right \rfloor} \left ( \frac{1}{N}\sum_{n=0}^{N-1}e^{-j2\pi n (\nu-\nu^r_{pq}-\frac{k\Delta f}{N})T} \right ) \nonumber \\ & \hspace{-20mm} & \hspace{-20mm} \left ( \frac{1}{M}\sum_{m=0}^{M-1}e^{j2\pi m \Delta f (\tau-\tau^r_{pq}-\frac{lT}{M})} \right ) \Bigg].
\label{zak7}
\end{eqnarray}
In converting the TD signal to DD domain in the Zak receiver, the DD domain signal is obtained by sampling the Zak transform of $y^r(t)$ at discrete points $(\tau=\frac{l'T}{M},\nu=\frac{k'\Delta f}{N})$, for $l'=0,\cdots, M-1$ and $k'=0,\cdots,N-1$. Therefore, the sampled DD domain signal in the Zak receiver is given by
\begin{eqnarray}
    y^r[k',l'] & \hspace{-2mm}\stackrel{\Delta }{=}\hspace{-2mm}& \mathcal{Z}_{y^r}\left(\tau=\frac{l'T}{M},\nu=\frac{k'\Delta f}{N}\right),\nonumber\\ & \hspace{-6mm} & \hspace{-4mm}=\sqrt T \sum_{n=0}^{N-1}y^r(nT+\frac{l'T}{M})e^{-j2 \pi n\frac{k'}{N}}.
    \label{zak8}
\end{eqnarray}
Substituting (\ref{zak7}) in (\ref{zak8}), we get
\begin{align}
    y^r[k',l']=\phi_r\sum_{k=0}^{N-1}\sum_{l=0}^{M-1}x[k,l]\hat{g}^r[k',l',k,l],
    \label{zak9}
\end{align}
where 
\begin{eqnarray}
\hat{g}^r[k',l',k,l] & \hspace{-2mm}=& \hspace{-2mm}\sum_{q=1}^{P_2}\sum_{p=1}^{P_1}\rho^r_{pq}g^r_qh^r_{p} e^{j2\pi \frac{\nu^r_{pq}}{\Delta f}(\frac{l'}{M}-\frac{\tau^r_{pq}}{T})} \nonumber \\ 
& \hspace{-2mm} & \hspace{-2mm} 
e^{j2\pi(\frac{k'}{N}-\frac{\nu^r_{pq}}{\Delta f}) \left \lfloor \frac{l'}{M}-\frac{\tau^r_{pq}}{T} \right \rfloor } \nonumber \\
& \hspace{-2mm} & \hspace{-2mm}
\left ( \frac{1}{N} \sum_{n=0}^{N-1}e^{-j2\pi nT(\frac{(k'-k)\Delta f}{N}-\nu^r_{pq})} \right )\nonumber \\ 
& \hspace{-2mm} & \hspace{-2mm} \left ( \frac{1}{M} \sum_{m=0}^{M-1}e^{j2\pi m \Delta f(\frac{(l'-l)T}{M}-\tau^r_{pq})} \right ), \nonumber
\end{eqnarray}
$k'=0,\cdots,N-1, l'=0,\cdots,M-1$.
Vectorizing (\ref{zak9}), we get
\begin{align}
    \mathbf{y}^r= \phi_r\mathbf{G}^r\mathbf{x},
    \label{zak10}
\end{align}
where $\mathbf{G}^r \in \mathbb{C}^{MN \times MN}$ is the effective cascaded channel matrix for the Zak receiver associated with the $r$th sub-surface with the element in its $(l'+k'M+1)$th row and $(l+kM+1)$th column being $\hat{g}^r[k',l',k,l]$. The overall input-output relation is obtained by adding the received signals $\mathbf{y}^r$ reflected from all the sub-surfaces at the receiver, as
\begin{eqnarray}
\mathbf{y}& \hspace{-2mm} = & \hspace{-2mm} \sum_{r=1}^{L}\phi_r \mathbf{G}^r \mathbf{x}+\mathbf{n},
\ = \ \mathbf{G}_{\text{eff}}\mathbf{x}+\mathbf{n},
\label{zak11}
\end{eqnarray}
where $\mathbf{G}_{\text{eff}}=\sum_{r=1}^{L}\phi_r \mathbf{G}^r$, $\mathbf{y}$ is the combined received vector from all the sub-surfaces and $\mathbf{n}$ is the additive noise vector.

{\em Complexity of single-step Zak receiver and two-step receiver:}
In (\ref{zak8}), for a given $l'$ and for all $k'=0,\cdots,N-1$, $y^{r}[k',l']$ can be computed using $N$-point discrete Fourier transform (DFT) and its complexity is $O(N\log N)$. Hence, the complexity of the Zak receiver for $k'=0,\cdots,N-1$, $l'=0,\cdots,M-1$ is $O(MN \log N)$. For the two-step receiver in Sec. \ref{sec2}, the complexity of calculating SFFT is $O(MN \log (MN))$.

\section{Results and discussions}
\label{sec4}
In this section, we present the simulation results on the bit error rate (BER) performance of RIS-aided OTFS with two-step receiver and single-step Zak receiver. A carrier frequency ($f_c$) of $4$ GHz, a subcarrier spacing ($\Delta f$) of $3.75$ kHz, a maximum Doppler of 1.2 kHz, BPSK modulation, and MMSE detection are considered. The channel fade coefficients $h^r_p$s and $g^r_q$s are assumed to be i.i.d and distributed as complex Gaussian with zero mean and variance $1/P$, where $P$ denotes the number of DD channel paths in the corresponding link. The Doppler shift corresponding to $l$th tap is generated using Jakes's formula, i.e., $\nu_l=\nu_{\max}\cos(\psi_l)$, where $\nu_{\max}$ denotes the maximum Doppler shift and $\psi_l$ is uniformly distributed in $[-\pi,\pi]$. The delay corresponding to $l$th tap in each link is generated uniformly in $[0,\frac{MT_s}{2}]$, where $T_s=\frac{1}{M\Delta f}$. For OTFS without RIS, $P$ is taken to be 4 and the delay corresponding to $l$th tap is generated uniformly in $[0,MT_s]$ and Doppler is generated using Jakes's formula.

{\em Reflection phase design:} At the RIS, reflection phase vector $\mathbf{\Theta}=[\theta_1 \hspace{0.5mm} \theta_2 \hspace{0.5mm} \cdots \hspace{0.5mm} \theta_L]$ is chosen such that Frobenius norm of the effective end-to-end DD channel matrix of the RIS-aided OTFS system is maximized. Let $\gamma_r=1$, $\theta_r \in [-\pi,\pi]$, $r=1,\cdots,L$ and $\mathbf{\Theta}^i$ be the $i$th realization. A large number of such phase vectors are generated and that vector which gives the maximum Frobenius norm is chosen, i.e., choose $\mathbf{\Theta}^{i^*}$ where $i^*=\arg\max_i \big \{\| \sum_{r=1}^{L}e^{j\theta_r^i}\mathbf{H}^r\|^2 \big \}$. 

\begin{figure}
\includegraphics[width=9.5cm,height=6.75cm]{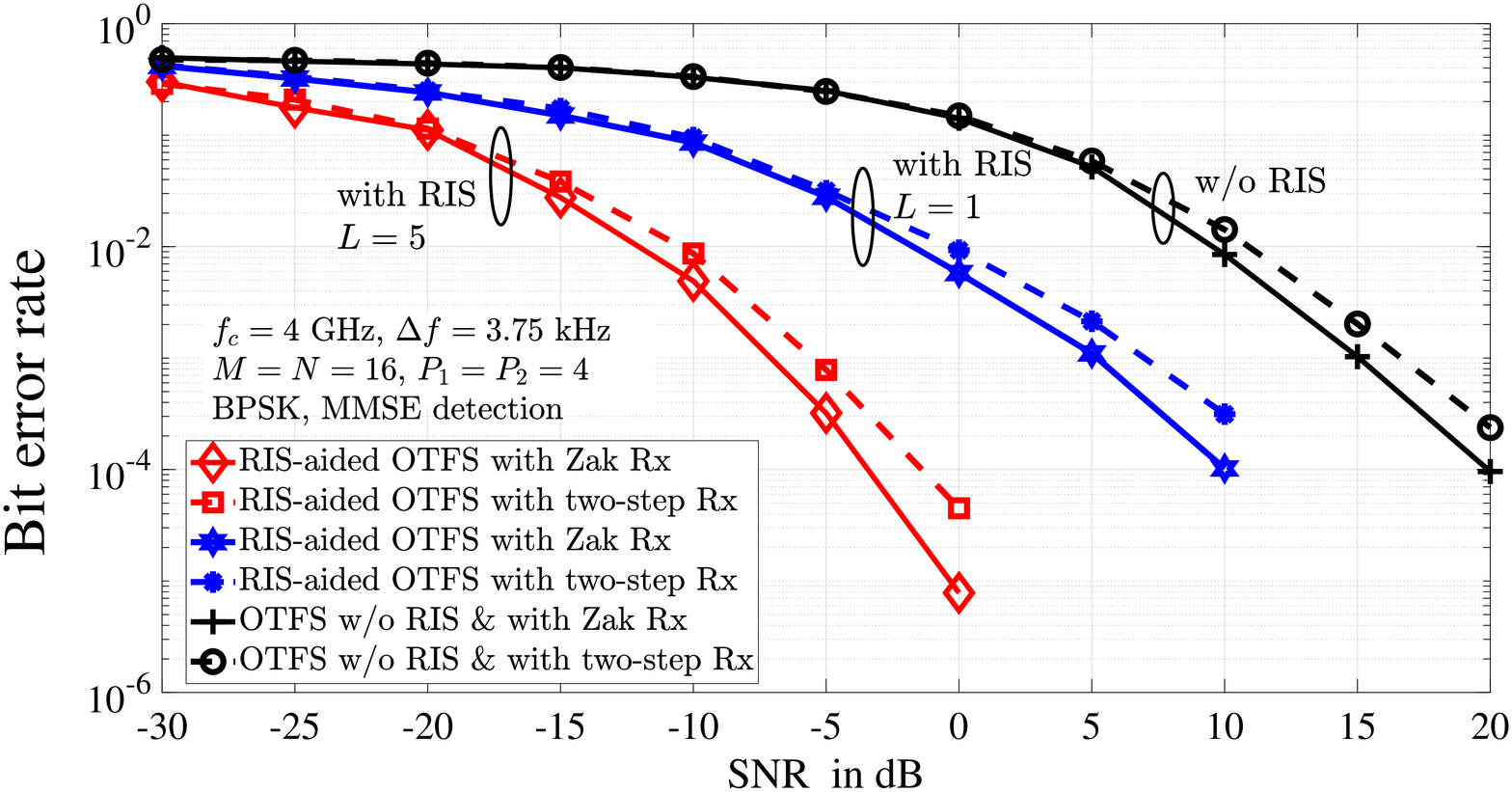}
\caption{BER performance of RIS-aided OTFS with  two-step and Zak receivers without RIS and with RIS.}
\vspace{-4mm}
\label{RIS_OTFS_2step_ZAk}
\end{figure}

{\em Performance with two-step receiver and Zak receiver:} Figure \ref{RIS_OTFS_2step_ZAk} shows the BER performance of RIS-aided OTFS for $M=N=16$ and $P_1=P_2=4$. We have plotted the BER for two-step receiver and Zak receiver for the following three cases: 1) without RIS, 2) with RIS for $L=1$, and 3) with RIS for $L=5$. SNR is defined as the ratio between the transmit power and noise variance, $\sigma^2$. Specifically, SNR is defined as $\frac{1}{\sigma^2}$, and the beamforming effect of the RIS will boost the signal power by a factor proportional to $L^2$, i.e., the SNR at the receiver is $\propto \frac{L^2}{\sigma^2}$ \cite{b4}. It is seen that RIS with OTFS offers significantly better performance compared to OTFS without RIS. This performance gain with RIS is due to the boost in the received SNR offered by reflections from the RIS sub-surfaces whose phases are tuned to enhance the SNR at the receiver. The effect of $L^2$ boost in SNR can be seen in Fig. \ref{RIS_OTFS_2step_ZAk} by comparing the performance of $L=1$ and $L=5$, where there is a SNR gap of about $10\log_{10}(5^2)=13.9$ dB at a BER of $10^{-3}$. It can also be seen that the Zak receiver performs better than the two-step receiver by about 2 dB at $10^{-3}$ BER.

\begin{figure}
\includegraphics[width=9.5cm,height=6.75cm]{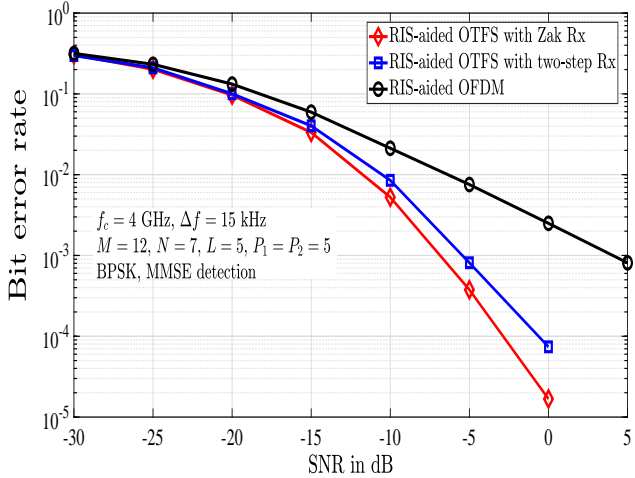}
\caption{BER performance comparison between RIS-aided OTFS and RIS-aided OFDM.}
\vspace{-4mm}
\label{RIS_OTFS_OFDM}
\end{figure}  

\begin{figure}[t]
\includegraphics[width=9.5cm,height=6.75cm]{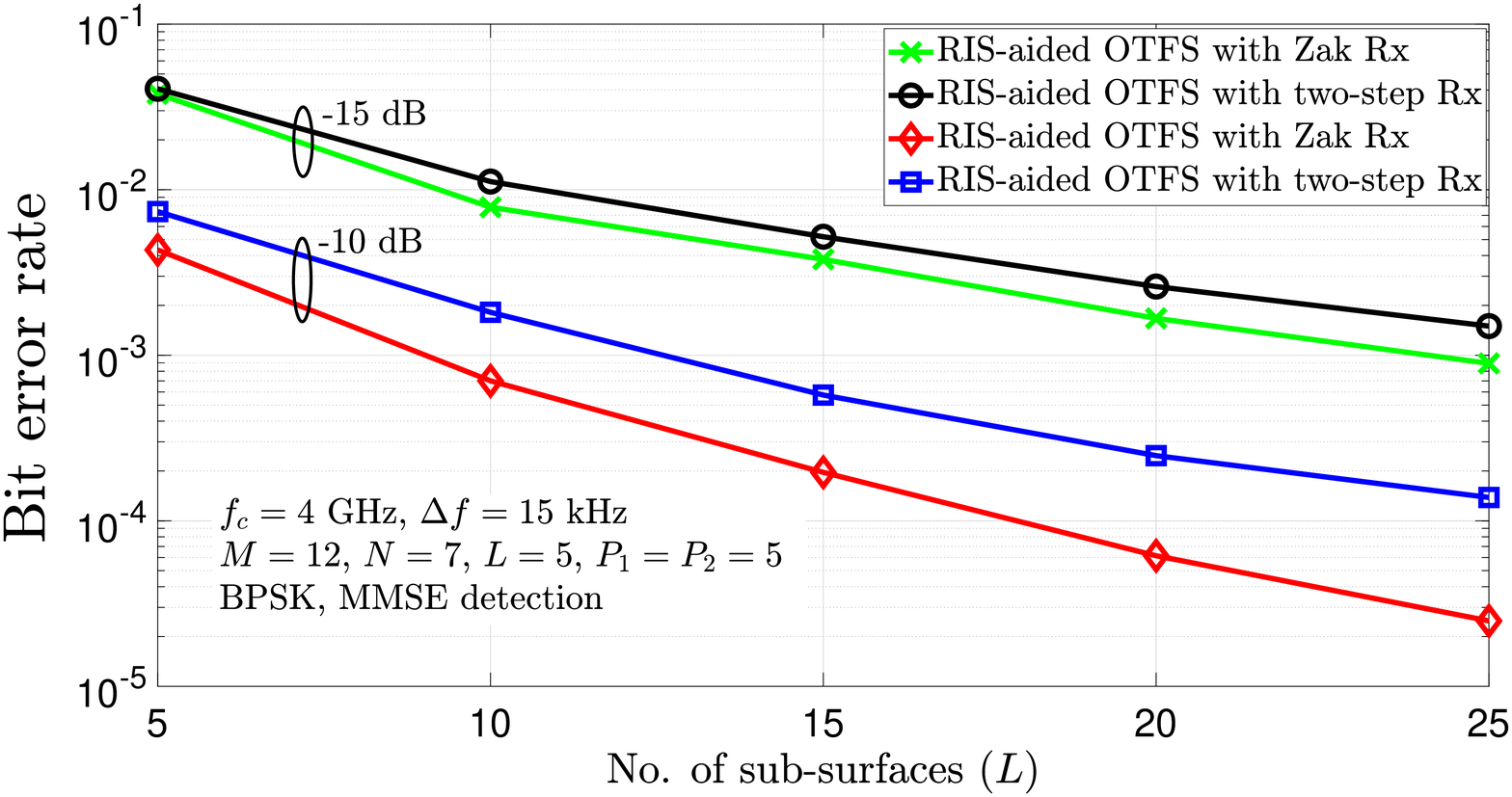}
\caption{Effect of number of sub-surfaces in the RIS on the BER performance of RIS-aided OTFS.}
\vspace{-4mm}
\label{RIS_OTFS_2step_ZAK_varyingL}
\end{figure} 

\begin{figure}
\includegraphics[width=9.5cm,height=6.75cm]{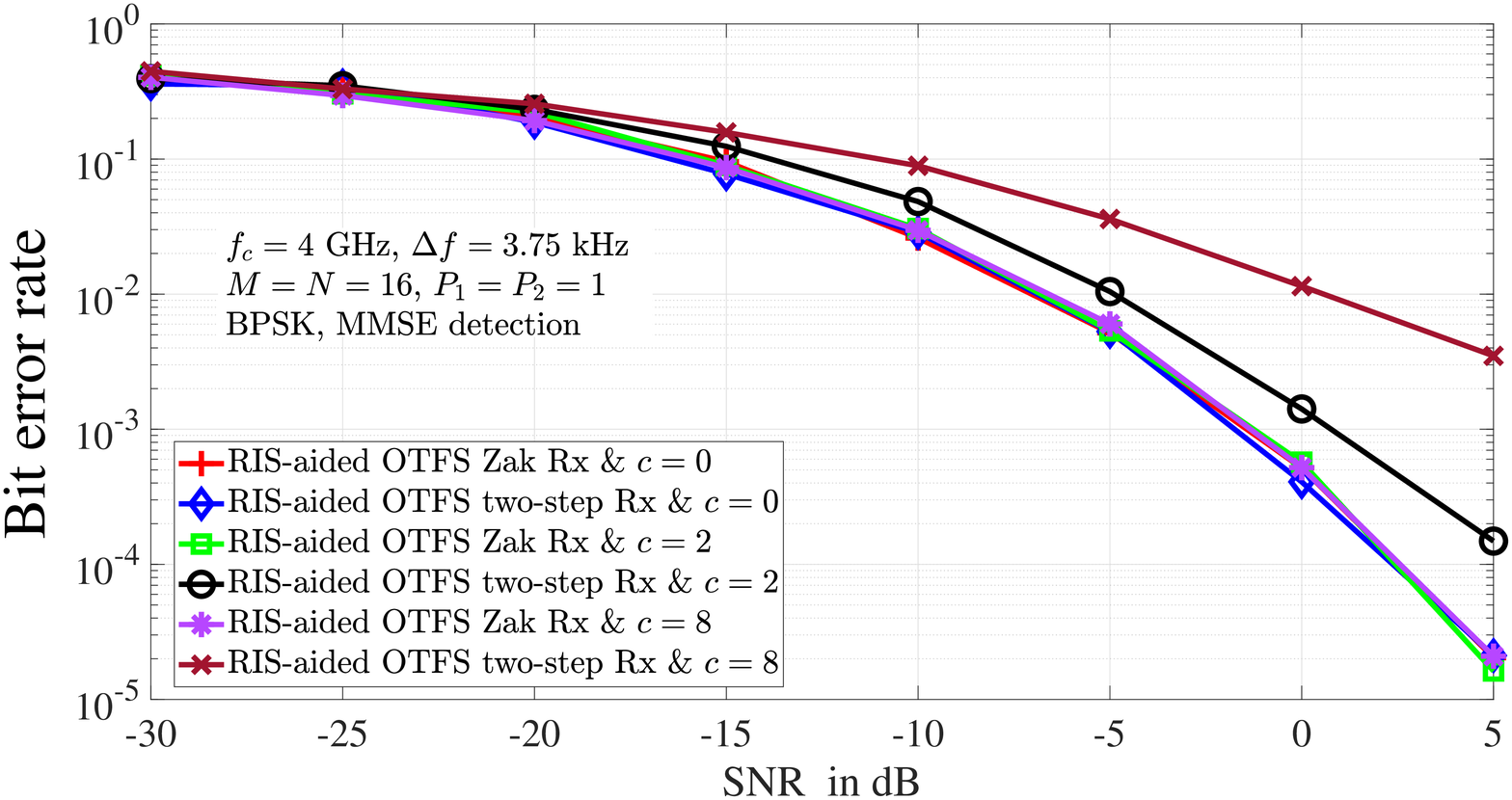}
\caption{BER performance comparison between Zak receiver and two-step receiver in RIS-aided OTFS.}
\vspace{-4mm}
\label{RIS_two_vs_zak}
\end{figure}

{\em Performance comparison between RIS-aided OTFS and RIS-aided OFDM:} Figure \ref{RIS_OTFS_OFDM} shows the performance comparison between RIS-aided OTFS and RIS-aided OFDM for $M=12$, $N=7$, which corresponds to the smallest resource block in LTE. A carrier frequency of 4 GHz, subcarrier spacing of 15 kHz, maximum Doppler shift of 1.85 kHz, and Jakes Doppler spectrum are considered.
The plots show that RIS-aided OTFS with Zak receiver achieves the best performance, followed by RIS-aided OTFS with two-step receiver. RIS-aided OFDM has the least performance. For example, at a BER of $10^{-3}$, RIS-aided OTFS with Zak receiver has an SNR gain of about 2 dB and 11 dB compared to the RIS-aided OTFS with two-step receiver and RIS-aided OFDM, respectively.

{\em Effect of number of sub-surfaces on performance:} Figure \ref{RIS_OTFS_2step_ZAK_varyingL} shows the effect of the number of sub-surfaces on the BER performance of RIS-aided OTFS with two-step receiver and Zak receiver for the parameters mentioned in Fig. \ref{RIS_OTFS_OFDM} at SNRs of -15 dB and -10 dB. It can be seen that, as expected, the BER improves as the number of sub-surfaces ($L$) increases. For example, at an SNR of -10 dB, the improvement in BER is two orders when $L$ is increased from 5 to 25. Also, increasing $L$ is observed to yield diminishing returns in BER performance, which can be explained as follows. For any two values of number of sub-surfaces $L_1$ and $L_2$, the performance gap is proportional to $\big(\frac{L_1}{L_2}\big)^2$, and the ratio $\frac{L_1}{L_2}$ diminishes as $L_1,L_2$ are increased keeping $L_1-L_2$ fixed, e.g., for $L_1=15$, $L_2=10$, $\frac{L_1}{L_2}=1.5$ and for $L_1=25$, $L_2=20$, $\frac{L_1}{L_2}=1.25$, where $L_1-L_2=5$ in both cases.
Also, the Zak receiver is found to perform better than the two-step receiver.

{\em Superior performance of Zak receiver over two-step receiver:} 
Here, we provide an explanation for the better performance of the Zak receiver over the two-step receiver. For this, consider the case of zero delay ($\tau^r_{pq}=0$) and non-zero Doppler with $\nu^r_{pq}=c \Delta f$, $c\in \{0,1,\cdots,M-1\}$. From (\ref{y_DD2}), the effective channel gain terms for the two-step receiver become $\hat{h}^r_2[k',l',k,l]=0$ and
$\hat{h}^r_1[k',l',k,l]=h^r_1g_1^re^{j2\pi \frac{cl'}{M}}\delta[k-k']\left (  \frac{1}{M}\sum_{m=0}^{M-1-c}e^{j2\pi m \frac{(l'-l)}{M}}\right )$, where $\delta[n]$ is the Kronecker delta function. Likewise, from (\ref{zak9}), the effective channel gain term for the Zak receiver becomes $\hat{g}^r[k',l',k,l]=h^r_1g_1^re^{j2\pi \frac{cl'}{M}}\delta[k-k']\delta[l-l']$.
It is clear that $\hat{h}^r_1[k',l',k,l]$ and $\hat{g}^r[k',l',k,l]$ are equal when $c=0$ (i.e., zero-Doppler case). Whereas, they are not same when $c\neq0$, i.e.,
$\left(\frac{1}{M}\sum_{m=0}^{M-1-c}e^{j2\pi m \frac{(l'-l)}{M}}\right )$ is not equal to the delta function $\delta[l-l']$. This means, while there is no leakage to other bins in the case of Zak receiver, there is leakage into other bins in the case of two-step receiver. This leakage contributes to poorer performance of the two-step receiver. This point is illustrated through simulations for different values of Doppler, $\nu^r_{pq}=c\Delta f$ for $c=0,2,8$, in Fig. \ref{RIS_two_vs_zak}. The BER plots in Fig. \ref{RIS_two_vs_zak} show that the performance of Zak and two-step receivers are the same when $c=0$. But, for $c=2,8$, the performance of two-step receiver degrades because of the leakage into other bins. This leakage increases for increasing Doppler and hence the performance with $c=8$ is worse than that with $c=2$. Whereas, Zak receiver retains its performance close to its performance with $c=0$. This shows Zak receiver to be more resilient to Doppler than two-step receiver. Finally, some key challenges in RIS-aided OTFS that can be taken up for future work include DD channel estimation, development of efficient RIS phase optimization techniques, and proof-of-concept implementations and testbeds. The effect of correlation among sub-surfaces on performance can also be investigated.

\section{Conclusions}
\label{sec5}
We investigated the performance benefits of using RIS in OTFS modulated systems in high-Doppler channels. We derived the end-to-end DD domain input-output relation in RIS-aided OTFS systems with rectangular pulses and fractional delays and Dopplers. The derived input-output relation can aid further investigations in RIS-aided OTFS relating to transceiver techniques/algorithms, performance evaluation, and implementation. Our results showed that RIS-aided OTFS achieves significantly better performance compared to RIS-aided OFDM in high-Doppler channels.

\end{document}